\documentclass[a4paper,11pt]{article}
\pdfoutput=1

\usepackage{jheppub1}
\usepackage[T1]{fontenc}
\usepackage{amssymb}
\usepackage{graphicx}
\usepackage{amsmath}
\usepackage{hyperref}
\usepackage{tensor}
\usepackage{epstopdf}
\usepackage{extarrows}
\newcommand{\td}{\text{d}}

\def\D {\mathcal{D}}

\def\Fi {\text{Fi}}

\def\Tr {{\text{Tr}}}

\begin{document}
\title{Gravity dualities of quantum distances}
%\title{\boldmath How large is a cross-section inside black holes?}
\author{Run-Qiu Yang}
\emailAdd{aqiu@tju.edu.cn}
%\email{aqiu@tju.edu.cn}

\affiliation{Center for Joint Quantum Studies and Department of Physics, School of Science, Tianjin University, Yaguan Road 135, Jinnan District, 300350 Tianjin, P.~R.~China}

%\begin{abstract}
\abstract{
By choosing modular ground state as the reference state, this paper finds that three most frequently-used distances and a quantum quasi-distance, i.e. the trace distance, Fubini-Study distance, Bures distance and R\'{e}nyi relative entropy, all have gravity dualities. Their gravity dualities have two equivalent descriptions: one is given by the integration of the area of a cosmic brane, the other one is given by the Euclidian on-shell action of dual theory and the area of the cosmic brane. It then applies these dualities into the 2-dimensional conformal field theory as examples and finds the results match with the computations of field theory exactly.
}
%\end{abstract}
%%%%%%%%%%%%%%%%%%%%%%%%%%%%%%%%%%%%%%
\maketitle
%\tableofcontents
\flushbottom

%%%%%%%%%%%%%%%%%%%%%%%%%%%%%%%%%%%%%%
\noindent

\section{Introduction}
%\emph{Introduction-.}
In recent years it has been suggested that quantum information theory and gravity theory have deep connection. The gauge/gravity duality, which shows an equivalence between strongly coupled quantum field theories (QFTs) and weakly coupled gravitational theories in one higher dimensions~\cite{Maldacena:1997re,Gubser:1998bc,Witten:1998qj}, offers us a powerful tool towards such connection. As a result, the quantum information theoretic considerations have provided various useful viewpoints in the studies of gauge/gravity duality and quantum gravity. One example is the Ryu-Takayanagi (RT) formula~\cite{Ryu:2006bv,Hubeny:2007xt,Chen:2019lcd}, which connects the area of a codimension-2 minimal surface in the dual spacetime and the entanglement entropy of the boundary QFT. The RT formula has been generalized into the R\'{e}nyi entropy~\cite{Hung:2011nu,Dong:2016fnf}, higher order gravity theory~\cite{Dong:2013qoa,Camps:2013zua,Miao:2014nxa} and the cases with quantum corrections~\cite{Faulkner:2013ana,Engelhardt:2014gca}. An other quantity in quantum information named ``complexity'', which measures the difference of two states according to the size of quantum circuits in converting one state into the other, also has been studied widely in gravity and black hole physics~\cite{Harlow:2013tf,Stanford:2014jda,Susskind:2014rva,Brown:2015bva,Czech:2017ryf,Caputa:2018kdj}.

From a general viewpoint, the complexity is a kind of ``distance'' between quantum states~\cite{Susskind:2014jwa}. Except for complexity, there are other several different measures of the distance between states, which are widely used in quantum information~\cite{Nielsen:2011:QCQ:1972505,Watrous}. For example, given two density matrices $\rho$ and $\sigma$ in the same Hilbert space, two families of distance are widely used in quantum information theory. The first one are based on the fidelity
\begin{equation}\label{deffidelity}
  \Fi(\rho,\sigma)=\Tr\sqrt{\sqrt{\sigma}\rho\sqrt{\sigma}}\,.
\end{equation}
The fidelity is not a distance but we can use it to define two kinds of distance, the Fubini-Study distance $D_F(\rho,\sigma)=\arccos\Fi(\rho,\sigma)$ and the Bures distance $D_B(\rho,\sigma)=\sqrt{1-\Fi(\rho,\sigma)}$. The other family of distances, depending on a positive number $n$, is provided by the $n$-distances
\begin{equation}\label{deftraceD}
  D_n(\rho,\sigma):=\frac1{2^{1/n}}(\Tr|\rho-\sigma|^n)^{1/n}\,.
\end{equation}
Here $\Tr|X|^n:=\sum_{i}\lambda_i^n$ and $\lambda_i$ is the $i$-th eigenvalue of $\sqrt{X^\dagger X}$. When $X$ is hermitian $\{\lambda_i\}$ are just the the absolute values of eigenvalues of $X$. Two special choices are widely applied.  One is Hilbert-Schmidt distance, which chooses $n=2$. This distance leads to some conveniences in mathematics because the calculation is straightforward by its definition. The other choice is $n=1$, which is called the ``trace distance''.

Though the trace distance and fidelity are complicated than Hilbert-Schmidt distance,  several properties make them special~\cite{Nielsen:2011:QCQ:1972505}. Firstly, the trace distance and fidelity (and so Fubini-Study distance and Bures distance) are bounded by each others: $1-D_1\leq\Fi\leq\sqrt{1-D_1^2}$. Secondly, they offer us dimension-independent bounds on the difference between the expected values of an operator $O$ in different states: $|\langle O\rangle_\rho-\langle O\rangle_\sigma|\leq D_1(\rho,\sigma)\sqrt{\Tr(OO^\dagger)}\leq \sqrt{1-\Fi(\rho,\sigma)^2}\sqrt{\Tr(OO^\dagger)}$. Thirdly, they supply lower bounds for the relative entropy $S(\rho\parallel\sigma)$: $[1-\Fi(\rho,\sigma)]\leq D_1(\rho,\sigma)\leq \sqrt{S(\rho\parallel\sigma)/2}$.

Despite that trace distance and fidelity have these important properties, their computations are high challenge in quantum field theory. The first breakthrough towards this issue is achieved by Ref.~\cite{PhysRevLett.113.051602}, which develops a replica trick to compute the fidelity for 2-dimensional conformal field theory. Refs.~\cite{Zhang:2019wqo,Zhang:2019itb,Zhang:2019kwu} then also develop replica method to compute the trace distance for a class of special states for single short interval in 1+1 dimensional CFTs. However, there are still huge difficulties in the calculations of trace distance even for 1+1 dimensional CFTs, such as to compute the trace distance between two thermal states or two large intervals in CFTs. There is also no compact method to compute the trace distance in higher dimensional CFTs or general quantum field theories. On the other hand, the holographic descriptions of entanglement, relative entropy~\cite{Blanco:2013joa} and complexity have been found, however, the trace distance does not yet. Refs.~\cite{PhysRevLett.115.261602,PhysRevD.96.086004} propose holographic duality to compute the fidelity between a state and its infinitesimal perturbational state, however, they cannot be used into the case when the difference between two states are not infinitesimal.

In this paper, it will develop holographic dualities to compute the trace distance, Fubini-Study distance, Bures distance and R\'{e}nyi relative entropy. By choosing a characteristic reference state which will be called ``modular ground state'', they all become the intrinsic properties of the target state. Then this paper will show that they all have gravity dualities. Their gravity dualities have two equivalent descriptions. The one is given by the integration of the area of a cosmic brane with respective to its tansion. The other one description contains two parts,  one of which is the on-shell action of gravity theory and the other one of which is the area of the cosmic brane. We then apply them to the calculations of the trace distance in 1+1 dimensional CFT and show our holographic calculations exactly match with the results of CFT's.

\section{Distances to modular ground state and holographic proposals}
%\emph{Distances to modular ground state and holographic proposals-.}
In the field theory two different density matrices will often be almost orthogonal $\Tr(\rho\sigma)\approx0$ and so their trace distance will almost saturate the upper bound. In this case, it will be more convenient to study a ``refined trace distance''
\begin{equation}\label{defrefined1}
  \D_T(\rho,\sigma)=-\ln[1-D_1(\rho,\sigma)]\,.
\end{equation}
Due to the monotonicity, the refined trace distance and trace distance contain same information. Similarly, we also defined a ``refined Fubini-Study distance'' $\D_F(\rho,\sigma)$ and ``refined Bures distance'' $\D_B(\rho,\sigma)$ as follows:
\begin{equation}\label{defrefined2}
  \D_F(\rho,\sigma)=-\ln\cos D_F(\rho,\sigma)\,,
\end{equation}
and
\begin{equation}\label{defrefined2}
  \D_B(\rho,\sigma)=-\ln[1-D_B(\rho,\sigma)^2]\,.
\end{equation}
One can verify that $\D_B(\rho,\sigma)=\D_F(\rho,\sigma)=-\ln\Fi(\rho,\sigma)$.

Differing from entanglement entropy which is an intrinsic property of target state $\rho$, the above three quantum distances do not only depend on the target state $\rho$ but also depend on a reference state $\sigma$. We can choose a characteristic reference state which belongs to the target state so that these quantum distances also become intrinsic properties of target states. If $\rho$ is a thermal state, one natural characteristic reference state is just the ground state. Such choice has been widely used in studying quantum phase transitions, e.g. see Refs.~\cite{PhysRevB.93.235160,Zanardi2007}. This choice can be generalized into arbitrary target states. Assume that $\rho$ is an arbitrary quantum state. As $\rho$ is both hermitian and positive semi-definite, we can formally define $\rho=e^{-K}$ with a hermitian operator $K$. Here $K$ is known as the modular Hamiltonian in axiomatic quantum field theory~\cite{9780387536101} or entanglement Hamiltonian in some literatures studying entanglement entropy~\cite{PhysRevLett.101.010504,PhysRevLett.105.080501}. To discuss the quantum distance for state $\rho$, we choose the reference state to be
\begin{equation}\label{defmod2}
  \Omega(\rho)=\lim_{n\rightarrow\infty}\Omega_n(\rho),~~\Omega_n(\rho):=\frac{\rho^n}{\Tr(\rho^n)}\,.
\end{equation}
We call this special reference state to be ``modular ground state'' of state $\rho$, as it is the ground state of modular Hamiltonian $K$. The quantum distance between $\rho$ and its modular ground state becomes an intrinsic quantity of quantum state $\rho$. We call such quantum distance to be ``intrinsic quantum distance'' of $\rho$. If the state $\rho$ is just a thermal state, then the modular ground state is just the zero temperature state of the system. We denote the ``intrinsic refined quantum distances'' to be
\begin{equation}\label{defDTFB1}
  \hat{\D}_T(\rho):=\D_T(\rho, \Omega(\rho)), ~~\hat{\D}_F(\rho):=\D_F(\rho, \Omega(\rho))
\end{equation}
and
\begin{equation}\label{defDFTB2}
  \hat{\D}_B(\rho):=\D_B(\rho, \Omega(\rho))\,.
\end{equation}
Our main results are holographic formulas for above three intrinsic refined quantum distances. They are given by two kinds of descriptions.

Assume that quantum state $\rho$ is dual to a boundary region $\mathcal{A}$ in a time slice of an asymptotically AdS spacetime. In the first description, the intrinsic refined quantum distances of state $\rho$ are related to the area in Planck units of a bulk codimension-2 cosmic brane $C_n$ which is homologous to the region $\mathcal{A}$:
\begin{equation}\label{hologreq2b}
  \hat{\D}_T(\rho)=2\hat{\D}_F(\rho)=\int_1^\infty\frac{\text{Area}(C_n)}{4G_Nn^2}\td n-\frac{\text{Area}(C_\infty)}{4G_N}\,.
\end{equation}
Here $G_N$ is the Newton's constant and we use the subscript $n$ on the cosmic brane to denote that its brane tension as a function of $n$ is given by $ T_n=(n-1)/(4n G_N)$. The geometry of above asymptotically AdS spacetime with above cosmic brane can be regarded as the solution an Euclidean gravity theory with the total action
\begin{equation}\label{actiong1}
  I_{\text{total}}^{(n)}=I_{\text{bulk}}+I_{\text{brane}}^{(n)}\,.
\end{equation}
Here $I_{\text{bulk}}=I_{\text{gravity}}+I_{\text{matters}}$, $I_{\text{gravity}}$ is the Euclidean Hilbert-Einstein action with negative cosmological constant, $I_{\text{matters}}$ is the action of matter fields, $I_{\text{brane}}^{(n)}=T_n\text{Area}(C_n)$. One can obtain the classical solution for theory~\eqref{actiong1} by minimizing total action for a given boundary subregion $\mathcal{A}$ and tension $T_n$. Due to the nonzero tension, the brane will backreact on the bulk geometry. Thus, the bulk geometry and position of cosmic $C_n$ depend on subscript $n$.

In the second description, we take $\mathcal{M}_n(\rho)$ to be the bulk domina of which the equal Euclidean time hypersurface $\Sigma$ satisfies $\Sigma=\mathcal{A}\cup C_n$. The intrinsic refined quantum distances of state $\rho$ are given by on-shell action of gravity theory and the area of cosmic brane in following way:
\begin{equation}\label{holotrace1}
  \hat{\D}_T(\rho)=2\hat{\D}_F(\rho)=I_{\text{bulk}}[\mathcal{M}_\infty]-I_{\text{bulk}}[\mathcal{M}_1]-I_{\text{brane}}^{(\infty)}\,.
\end{equation}
By Eqs.~\eqref{holotrace1} and \eqref{hologreq2b}, the calculations of intrinsic refined quantum distances become solving partial differential equations in an Euclidean gravity theory with cosmic brane. When the variation of $I_{\text{total}}$ admits more than one classical solutions, we have to choose the one which has the smallest bulk action $I_{\text{bulk}}$ rather than the one which has smallest total action~\eqref{actiong1}. The reason is similar to the discussion in Ref.~\cite{Dong:2016fnf} and will be clarified briefly later.
%~\cite{Dong:2016fnf}.
%The holographic duality~\eqref{hologreq2} contains two parts: one contains the on-shell action of gravity theory and other other has a form qualitatively similar to the RT formula.

%The holographic dualities of $\D_T(\rho), \D_F(\rho)$ and $\D_B$ are give by the difference of on-shell actions when tension is zero tension and maximal.
\section{Deviation via holographic replica trick}
%\emph{Deviation via holographic replica trick-.}
We now present the holographic deviations on Eqs.~\eqref{holotrace1} and \eqref{hologreq2b}. We first consider intrinsic refined trace distance. The even order $n$-distance between $\rho$ and $\Omega_m(\rho)$ satisfies
\begin{equation}\label{general1}
  D_{2n}(\rho,\Omega_m)^{2n}=\frac1{2}\sum_{k=0}^{2n}C_{2n}^k(-1)^k\frac{\Tr(\rho^{k+m(2n-k)})}{\Tr(\rho)^k\Tr(\rho^m)^{2n-k}}\,.
\end{equation}
Here $C_{2n}^k:=(2n)!/[k!(2n-k)!]$ is the combinatorial number and $\Omega_m$ is defined by Eq.~\eqref{defmod2}. For convenience, we here do not assume $\rho$ is normalized. The intrinsic refined trace distance then is obtained by the limit $m\rightarrow\infty$ and analytical continuation $n=1/2$. Define $\exp(-\mathcal{F}_{m,n,k}(\rho))=\Tr(\rho^{k+m(2n-k)})/[\Tr(\rho)^k\Tr(\rho^m)^{2n-k}]$ and we will have
\begin{equation}\label{definF1}
  \mathcal{F}_{m,n,k}=(2n-k)\ln\Tr(\rho^m)+k\ln\Tr(\rho)-\ln\Tr(\rho^{k+m(2n-k)})\,.
\end{equation}

Now we use the gravity replica method~\cite{Faulkner:2013ana,Dong:2016fnf} to compute the $\mathcal{F}_{m,n,k}$. Follows the usual holographic dictionary, the trace of $\rho^m$ is given by the partition function of the QFT on a branched cover $M_m$.  Here $M_m$ is $m$-fold cover branched over $\mathcal{A}$, which is defined by taking $m$ copies of the original Euclidean spacetime $M_1$ where the QFT lives with a cut along the entangling region and gluing them along the cuts in a cyclic order.  In the large $N$ limit, the bulk physics is classical and we have $\Tr(\rho^m)=e^{-I_\text{bulk}(M_m)}$. The branched cover $M_m$ has a manifest $\mathbb{Z}_m$ symmetry, which is not spontaneously broken in the dominant bulk solution~\cite{Lewkowycz2013}. Take this  $\mathbb{Z}_m$ replica symmetry into account and we can define an orbifold $\mathcal{M}_m:=M_m/\mathbb{Z}_m$. As a result, we the bulk on-shell action becomes $I_\text{bulk}(M_m)=mI_\text{bulk}(\mathcal{M}_m)$. Then we obtain
\begin{equation}\label{definF2}
\begin{split}
  \mathcal{F}_{m,n,k}=&[k+(2n-k)m]I_\text{bulk}(\mathcal{M}_{k+(2n-k)m})\\
  &-(2n-k)mI_\text{bulk}(\mathcal{M}_m)-kI_\text{bulk}(\mathcal{M}_1)\,.
  \end{split}
\end{equation}
After the quotient of $\mathbb{Z}_m$, there are conical singularities which are the fixed point of $\mathbb{Z}_m$ quotient.  Such conical singularities are some co-dimensional 2 surfaces. The cosmic brane is added into the total action~\eqref{actiong1} so that the variation problem can just generate the solutions with such conical singularities. The real action of theory in the branched cover $M_m$ only contains the bulk term. This explains why the we need to choose the classical solution minimizing $I_{\text{bulk}}$ when theory~\eqref{actiong1} has multiple classical solutions.

Though the parameters $m,n$ and $k$ are assumed to be integers in the definition of $\mathcal{F}_{m,n,k}$, we will analytically continues them into nonnegative real numbers. For large $m$, we have two cases: if $k\neq2n$ we have
\begin{equation}\label{definF3}
\begin{split}
  \mathcal{F}_{m,n,k}&=k[I_\text{bulk}(\mathcal{M}_{\infty})-I_\text{bulk}(\mathcal{M}_1)-m^2\partial_mI_\text{bulk}(\mathcal{M}_m)]\\
  &+(2n-1)m^2\partial_mI_\text{bulk}(\mathcal{M}_m)+\mathcal{O}(1/m)\,
  \end{split}
\end{equation}
 and if $k=2n$ we have
\begin{equation}\label{definF3b}
  \mathcal{F}_{m,n,2n}=2n[I_\text{bulk}(\mathcal{M}_{2n})-I_\text{bulk}(\mathcal{M}_{1})]
\end{equation}
On the other hand, it has been shown~\cite{Dong:2016fnf}
\begin{equation}\label{gravitySq2}
  m^2\partial_mI_\text{bulk}(\mathcal{M}_m)=\frac{\text{Area}(C_m)}{4G_N}\,.
\end{equation}
Using Eqs.~\eqref{definF3} and \eqref{definF3b}, we find that the summation~\eqref{general1} can be computed analytically. Thus we find (see appendix~\ref{app1} for mathematical details)
\begin{equation}\label{general2}
  D_{1}(\rho,\Omega_\infty)=1-e^{-\mathcal{F}_0}\Rightarrow\D_T(\rho)=\mathcal{F}_0\,.
\end{equation}
where
\begin{equation}\label{valyefmnk1}
  \mathcal{F}_{0}:=I_\text{bulk}(\mathcal{M}_{\infty})-I_\text{bulk}(\mathcal{M}_1)-I^{(\infty)}_{\text{brane}}\,.
\end{equation}
Thus, we obtain the holographic formula~\eqref{holotrace1} for intrinsic refined trace distance. The holographic formula~\eqref{hologreq2b} can be obtained after we integrate Eq.~\eqref{gravitySq2} and use the fact $I^{(\infty)}_{\text{brane}}=\text{Area}(C_\infty)/(4G_N)$.

%%
%
%%
%
%
%
%%\section{Expressions in terms of R\'{e}nyi entropy}
%In order to obtain our holographic formula~\eqref{hologreq2b}, let us define auxiliary function
%%
%\begin{equation}\label{defSn}
%  \Sr_m:=\frac{m}{m-1}[I_\text{bulk}(\mathcal{M}_{m})-I_\text{bulk}(\mathcal{M}_1)]\,.
%\end{equation}
%%
%Ref.~\cite{Dong:2016fnf} uses holographic replica trick to prove that $\Sr_m$ is just the $m$-th R\'{e}nyi entropy of state $\rho$. It is easy to see $\mathcal{F}_{0}=\Sr_m-I^{(\infty)}_{\text{brane}}+\mathcal{O}(1/m)$. Expand $\Sr_m$ as the power series $\Sr_m=S_0+S_1/m+S_2/m^2\cdots$ and combine Eqs.~\eqref{defSn} and Eq.~\eqref{gravitySq2}. We then find $\text{Area}(C_m)/(4G_N)=S_0-S_1+2(S_1-S_2)/m+\cdots$ and so we see
%%
%\begin{equation}\label{valueFS1}
%  \mathcal{F}_{0}=S_1=-\frac{1}{8G_N}\lim_{m\rightarrow\infty}m^2\partial_m\text{Area}(C_m)\,.
%\end{equation}
%%
%Take it into Eq.~\eqref{general2} and we obtain the holographic formula~\eqref{hologreq2b} for intrinsic refined trace distance.

%\section{Other quantum distances}
In order to compute the fidelity, we consider a more general expression, i.e.  R\'{e}nyi relative entropies of two states, which in general is defined as follow~\cite{doi:10.1063/1.4838856,Wilde2014,PhysRevLett.113.051602}
\begin{equation}\label{relS1}
  S_k(\rho\|\sigma)=\frac1{k-1}\ln\Tr[(\sigma^{\frac{1-k}{2k}}\rho\sigma^{\frac{1-k}{2k}})^k]\,.
\end{equation}
The R\'{e}nyi relative entropy is just a quasi-distance function as it is not symmetric in general, i.e. $S_k(\rho\|\sigma)\neq S_k(\sigma\|\rho)$.
It is clear that $2\D_F(\rho,\sigma)=S_{1/2}(\rho\|\sigma)$. The limit $\lim_{k\rightarrow1}S_k(\rho\|\sigma)=S(\rho\|\sigma)$ is just he relative entropy between $\rho$ and $\sigma$. Take $\sigma=\Omega_m(\rho)$ and we find
\begin{equation}\label{relS2}
  S_k(\rho\|\Omega_m)=\frac1{k-1}\ln\Tr(\rho^k\Omega_m^{1-k})=\frac{\mathcal{F}_{m,1/2,k}}{1-k}\,.
\end{equation}
The intrinsic refined Fubini-Study distance and intrinsic refined Bures distance are obtained by
\begin{equation}\label{fubs1a}
  \hat{\D}_F(\rho)=\hat{\D}_B(\rho)=\frac12S_{1/2}(\rho\|\Omega)=\mathcal{F}_{\infty,1/2,1/2}=\mathcal{F}_{0}/2\,.
\end{equation}
Take Eq.~\eqref{valyefmnk1} into Eq.~\eqref{fubs1a} and we then obtain dualities of $\hat{\D}_F(\rho)$ and $\hat{\D}_B(\rho)$.

Using Eq.~\eqref{valyefmnk1}, we also find a holographic duality for the $k$-th R\'{e}nyi relative entropy between $\rho$ and its modular ground state when $k\in(0,1)$
\begin{equation}\label{holotrace1c}
\begin{split}
  S_{k}(\rho\|\Omega)&=\frac{k}{1-k}\left\{I_{\text{bulk}}[\mathcal{M}_\infty]-I_{\text{bulk}}[\mathcal{M}_1]-I_{\text{brane}}^{(\infty)}\right\}\\
  &=\frac{k}{1-k}\left[\int_1^\infty\frac{\text{Area}(C_n)}{4G_Nn^2}\td n-\frac{\text{Area}(C_\infty)}{4G_N}\right]\,.
  \end{split}
\end{equation}
The relative entropy $S(\rho\|\Omega)$ is given by the limit $k\rightarrow1$, which is divergent. This agrees with the analysis from quantum information theory.
%As a check of self-consistence, we consider the relative entropy between $\rho$ and its modular ground state $\Omega$.
By definition of relative entropy, we have $S(\rho\|\sigma)=\Tr(\rho\ln\rho)-\Tr(\rho\ln\sigma)$. When the state $\sigma=\Omega$, we can find that $\Tr(\rho\ln\sigma)=\infty$ and so $S(\rho\|\Omega)=\infty$. Note the R\'{e}nyi relative entropy is not symmetric about two states when $k\neq1/2$. However, there is a special permutation symmetry $(1-k)S_{k}(\Omega\|\rho)=kS_{1-k}(\rho\|\Omega)$. We then have
\begin{equation}\label{holotrace1c2}
\begin{split}
  S_k(\Omega\|\rho)&=I_{\text{bulk}}[\mathcal{M}_\infty]-I_{\text{bulk}}[\mathcal{M}_1]-I_{\text{brane}}^{(\infty)}\\
  &=\int_1^\infty\frac{\text{Area}(C_n)}{4G_Nn^2}\td n-\frac{\text{Area}(C_\infty)}{4G_N}\,.
  \end{split}
\end{equation}
Thus, we see $S_k(\Omega\|\rho)=\hat{\D}_T(\rho)$ is independent of $k$. As a self-consistent check, we can compute $S_k(\Omega\|\rho)$ directly in qubit system. Assume that $\rho$ is density matrix in a finite dimensional Hilbert space and $\lambda_0$ is its largest eigenvalue. Then we can find $S_k(\Omega\|\rho)=\frac1{k-1}\ln\Tr(W^k\rho^{1-k})=\frac1{k-1}\ln\Tr(W\rho^{1-k})=-\ln\lambda_0$, which is independent of $k$ as expected.
%%
%\begin{equation}\label{relsw1}
%  S(\rho\|\Omega)=\Tr(\rho\ln\rho)=-S_E(\rho)\,.
%\end{equation}
%%
%Here $S_E(\rho)$ is the entanglement entropy of $\rho$. In holographic side, the relative entropy is given by the limit $\lim_{k\rightarrow1}S_k(\rho\|\Omega)=\lim_{n\rightarrow1/2}\mathcal{F}_{\infty,n,2n}/(1-2n)$. Using Eqs.~\eqref{definF4} and \eqref{gravitySq2}, we find
%%
%\begin{equation}\label{relfn1}
%\begin{split}
%  S(\rho\|\Omega)&=\lim_{n\rightarrow1/2}\frac{2n[I_\text{bulk}(\mathcal{M}_{2n})-I_\text{bulk}(\mathcal{M}_{1})]}{1-2n}\\
%  &=-\lim_{n\rightarrow1}\partial_nI_\text{bulk}(\mathcal{M}_{n})=-\frac{\text{Area}(C_1)}{4G_N}\,.
%  \end{split}
%\end{equation}
%%
%This shows that the relative entropy between a state $\rho$ and its modular ground state is also given by RT surface corresponding to $\rho$. This agrees with Eq.~\eqref{relsw1}.

%This formula will bring convenience in some cases, however, the price is that we have to know the analytical continuation for area of brane as $n\rightarrow\infty$. Instead, to use Eq.~\eqref{hologreq2} we only need to find the solutions of tension $T_1=0$ and $T_{\infty}=1/(4G_N)$, which will be more convenient in numerical calculations.

%\section{holographic duality of in fidelity}

\section{Application in CFTs}
%\emph{Application in CFTs-.}
In following, we will show examples about how to use the holographic formula~\eqref{hologreq2b} to find the intrinsic trace distance of some states in CFT$_2$. In the first example, we consider a spherical disk $A$ with radius $R$ in a $d$-dimensional vacuum state. In princple, we need to solve Einstein's equation with the cosmic brane. However, as the disk is spherical, the task of finding the cosmic brane solution can be essentially simplified. In this case we can use conformal map of Ref.~\cite{Hung:2011nu} to direct obtain the bulk geometry. After the conformal map, the bulk geometry is a $d+1$-dimensional Euclidean hyperbolic AdS black hole~\cite{Dong:2016fnf,Nakaguchi:2016zqi}
\begin{equation}\label{adshyg1}
  \td s^2=\frac{\td \tau^2}{f_n(r)}+f_n(r)\td r^2+r^2[\td u^2+\sinh^2(u)\td\Omega_{d-2}^2]\,.
\end{equation}
and $f_n(r)=r^2-1-r_n^{d-2}(r_n^2-1)/r^{d-2}$. The cosmic brane is mapped into the horizon. The Euclidean time direction $\tau$ has the period $2\pi$ and so leads to the conical singularity at $r=r_n$. To match with the cosmic brane, we have to set $f'_n(r_n)=4\pi/n$. Then we find $r_n=[1+\sqrt{1+n^2d(d-2)}]/(nd)$
%%
%\begin{equation}\label{valuern}
%  r_n=[1+\sqrt{1+n^2d(d-2)}]/(nd)\,
%\end{equation}
%%
and $r_\infty=\sqrt{(d-2)/d}$. As the cosmic brane is just the horizon, we find Area$_n(C)=V_{d-1}(R)r_n^{d-1}$ with
\begin{equation}\label{valumH2}
  V_{d-1}(R)=\Omega_{d-2}\int_0^{\ln(2R/\epsilon)}\sinh^{(d-2)}(u)\td u\,.
\end{equation}
Here $\Omega_{d-2}=2\pi^{(d-1)/2}/\Gamma(\frac{d-1}2)$ is the area of the unit $(d-2)$-sphere. The upper limit of integration is the UV-cut off~\cite{Hung:2011nu}. In the case $d=2$, we find $r_n=1/n$ and Eq.~\eqref{hologreq2b} gives us a simple result
\begin{equation}\label{twodda1}
  \hat{\D}_T(\rho_A)=\ln(2R/\epsilon)/(4G_N)=\frac{c}6\ln(2R/\epsilon)\,.
\end{equation}
Here $c=3/(2G_N)$ is the central charge. In the appendix~\ref{app2} we will give a calculation from CFT side and show that two results match with each other exactly.
%We can compare this result with calculation of pure CFT. For a single interval of length $2R$, the leading term of R\'{e}nyi entropy in large $c$ limit reads $S_n=\frac{c}6(1+1/n)\ln(2R/\epsilon)$. Take this into Eq.~\eqref{hologreq1} and we find an exact match.
%

\begin{figure}[h!]
  \centering
  \includegraphics[width=.3\textwidth]{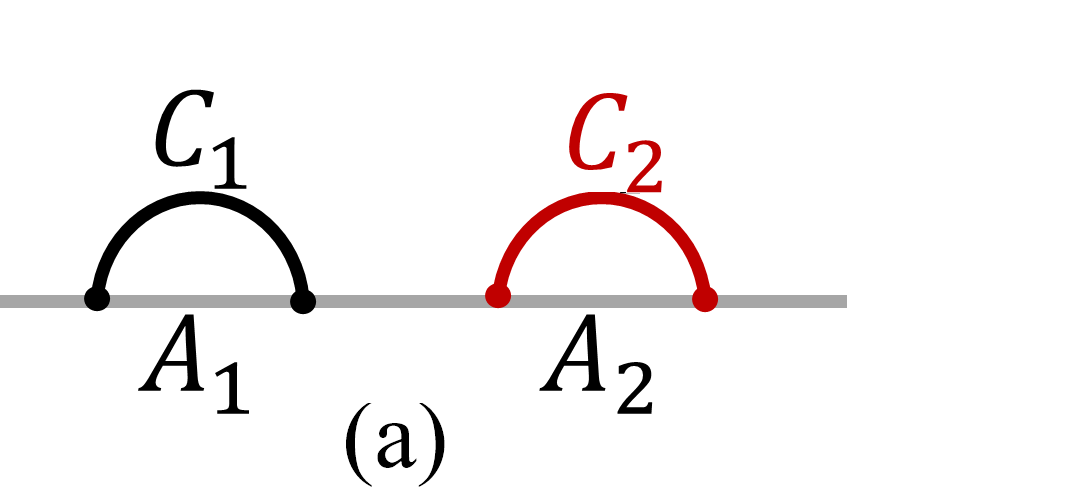}
  \includegraphics[width=.24\textwidth]{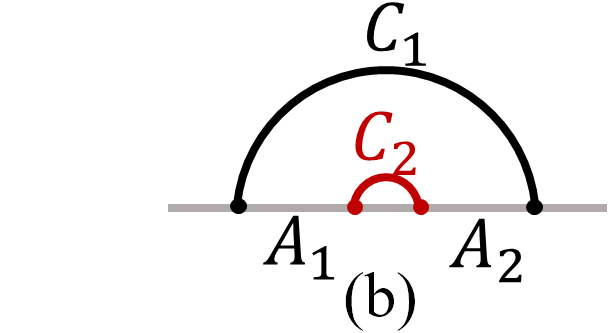}
  \caption{The cosmic branes of two intervals in the limit $x\ll1$ (subfigure (a)) and $x\rightarrow1$ (subfigure (b)).}
  %The possible inner horizons and singularities are irrelevant, so they are not showed.
   \label{diskfig}
\end{figure}

In the second example, we assume that the subregion $A$ contains two symmetric disjoint intervals in 2D case, i.e. $A=A_1\cup A_2$, where $A_1=[0,l]$ and $A_2=[1,1+l]$.  The cross ration $x=l^2$.  We first consider the limit $x\ll1$, which means two cosmic brane will be separated far enough. See the subfigure (a) of Fig.~\ref{diskfig}.
Though every brane will backreact on the bulk geometry, the interaction between two branes will be suppressed. Thus, up to the leading order of $x$, the final result would be simply twice of a single brane.  The intrinsic refined trace distance in this case becomes
\begin{equation}\label{twodda2}
  \hat{\D}_T(\rho_A)\approx\frac{c}6\ln(l/\epsilon)\times2=\frac{c}6\ln(x/\epsilon^2)\,.
\end{equation}
%
%The CFT's calculation shows that the R\'{e}nyi entropy at large $c$ and small $x$ limit reads reads~\cite{Headrick:2010zt,Hartman:2013mia,Perlmutter:2013paa,Chen:2013kpa}
%
%\begin{equation}\label{sntwoeq1}
%  S_n=\frac{c}3(1+1/n)\ln(l/\epsilon)+\mathcal{O}(x^2)\,.
%\end{equation}
%%
%Take it into Eq.~\eqref{hologreq1} and we obtain Eq.~\eqref{twodda2} exactly.
For large cross ration $x\rightarrow1$, the configuration of two cosmic branes is shown in subfigure (b) of Fig.~\ref{diskfig}. The second cosmic brane $C_2$ will shrink into the boundary and two brane will also decouple with each other. As the result, we can compute two branes separately and obtain
\begin{equation}\label{twodda2b}
  \hat{\D}_T(\rho_A)\approx\frac{c}6\{\ln[(1+l)/\epsilon]+\ln[(1-l)/\epsilon]\}=\frac{c}6\ln[(1-x)/\epsilon^2]\,.
\end{equation}
Comparing with Eq.~\eqref{twodda2}, we see that two limits have a symmetry $x\rightarrow1-x$. Note that the computation here only involves the leading terms of $x\rightarrow0$ or $x\rightarrow1$. It will be interesting to study what will happen if the interaction between two branes cannot be neglected in the future.

\section{Summary}
%\emph{Discussion-.}
To summary, this paper studies the holographic dualities of three most frequently-used quantum distances and a quantum quasi-distance, i.e. the trace distance, Fubini-Study distance, Bures distance and R\'{e}nyi relative entropy. By choosing the modular vacuum as the reference state, it finds that they all have holographic dualities. Then it applies these holographic dualities into 2-dimensional CFTs and show that holographic results exactly match with the calculations of field theory.

For the holographic formula \eqref{hologreq2b}, there is no difficulty to obtain generalizations to theories dual to higher derivative gravity. The basic idea is similar to the directions of Refs.~\cite{Dong:2013qoa,Camps:2013zua,Miao:2014nxa}. The area term in right-hand side of Eq.~\eqref{hologreq2b} should be replaced by the Wald entropy~\cite{Wald:1993nt} evaluated on a cosmic brane. In addition, following the methods of~\cite{Faulkner:2013ana,Engelhardt:2014gca} we can also include quantum corrections into Eq.~\eqref{holotrace1} by taking the bulk matters into account. For 2 dimensional CFT with 2+1 dimensional gravity duality, it is also interesting to consider the perturbational expansion of small cross ration in gravity side by taking the interaction of two cosmic branes and then compare it with the results of CFTs.

\acknowledgments
%\emph{Acknowledgments.--} %The authors would like to thank...
The work is supported by the Natural Science Foundation of China under Grant No. 12005155.

\appendix
\section{About Eq.~\eqref{general2}}\label{app1}
We first note that, for large $m$,
\begin{equation}\label{largemM1}
  I_\text{bulk}(\mathcal{M}_m)=I_\text{bulk}(\mathcal{M}_\infty)+I_1/m+\mathcal{O}(1/m^2)\,.
\end{equation}
If $k\neq2n$, then we also have
\begin{equation}\label{largemM2}
  I_\text{bulk}(\mathcal{M}_{k+(2n-k)m})=I_\text{bulk}(\mathcal{M}_\infty)+\frac{I_1}{k+(2n-k)m}+\mathcal{O}(1/m^2)\,.
\end{equation}
This gives us Eq.~\eqref{definF3} at large $m$ limit when $k\neq2n$.  
We then take Eqs.~\eqref{definF3} and \eqref{definF3b} into Eq.~\eqref{general1} and take the limit $m\rightarrow\infty$. Then we find
\begin{equation}\label{general1f1}
\begin{split}
  &D_{2n}(\rho,\Omega)^{2n}\\
  =&\frac1{2}\sum_{k=0}^{2n-1}C_{2n}^k(-1)^ke^{-k a-(2n-1)I_0}+e^{-B_{2n}}\,.
  \end{split}
\end{equation}
Here
$$a=I_\text{bulk}(\mathcal{M}_{\infty})-I_\text{bulk}(\mathcal{M}_1)-\frac{\text{Area}(C_\infty)}{4G_N},~~I_1=-m^2\partial_mI_\text{bulk}(\mathcal{M}_m)\,,$$
and
$$B_{2n}=-2n[I_\text{bulk}(\mathcal{M}_{2n})-I_\text{bulk}(\mathcal{M}_{1})]\,.$$
Then the summation~\eqref{general1f1} can be computed analytically
\begin{equation}\label{general1f2}
\begin{split}
  &D_{2n}(\rho,\Omega)^{2n}\\
=&\frac{e^{-(2n-1)I_1}}{2}\sum_{k=0}^{2n}C_{2n}^k(-1)^ke^{-k a}-e^{-2na}+e^{B_{2n}}\\
  =&\frac{e^{-(2n-1)I_1}}{2}\left[\left(1-e^{-a}\right)^{2n}-e^{-2na}+e^{B_{2n}}\right]\,.
  \end{split}
\end{equation}
Analytically continue it into $n=1/2$ and we find
\begin{equation}\label{general1f3}
  D_{1}(\rho,\Omega)=1-e^{-a}\,.
\end{equation}
Thus, we find
\begin{equation}\label{appdt1}
  \hat{\D}_T(\rho)=I_\text{bulk}(\mathcal{M}_{\infty})-I_\text{bulk}(\mathcal{M}_1)-I^{(\infty)}_{\text{brane}}\,.
\end{equation}
This gives us Eq.~\eqref{general2}.

\section{Calculations from CFT}\label{app2}
To calculate the intrinsic refined trace distance from CFT, we still start from the Eqs.~\eqref{general1} and \eqref{definF1}. The difference is that now we will use the replica trick of field theory. In 2D CFT, the trace of $\rho^m$ can be obtained by using twist operators. Follows the usual computations in CFT$_2$ (e.g. see Ref.~\cite{Chen:2019lcd}), one can find that the result in large $c$-limit reads
\begin{equation}\label{cfttracerm1}
 \ln\Tr(\rho^m)=-\frac{c}6(m-1/m)\ln(2R/\epsilon)\,.
\end{equation}
Taking this into Eq.~\eqref{definF1} and considering the large $m$ limit, we find
\begin{equation}\label{definF1f1}
  \mathcal{F}_{m,n,k}=-\frac{(2n-1)c}6\ln(2R/\epsilon)+\frac{kc}6\ln(2R/\epsilon)+\mathcal{O}(1/m)\,.
\end{equation}
if $k\neq2n$ and
\begin{equation}\label{definF3f1}
  \mathcal{F}_{m,n,2n}=\frac{c}6[2n-1/(2n)]\ln(2R/\epsilon)\,.
\end{equation}
Then we take Eqs.~\eqref{definF1f1} and \eqref{definF3f1} into Eq.~\eqref{general1}. Following the same steps of appendix~\ref{app1}, the summation can be computed analytically and we finally find that the trace distance reads
\begin{equation}\label{cftd1}
  D_1(\rho,\Omega)=1-\exp\left[-\frac{c}6\ln(2R/\epsilon)\right]\,.
\end{equation}
Thus, we see $\hat{\D}_T(\rho)=\frac{c}6\ln(2R/\epsilon)$, which matches with the our holographic result exactly.
%\begin{acknowledgments}
%The author would like to thank Jia-Ju Zhang for his helpful discussions.
%\end{acknowledgments}

\bibliographystyle{JHEP}
\bibliography{traceD}

\end{document}